\documentclass[aps,prd,showpacs,nofootinbib,tightenlines]{revtex4}
\usepackage{amsmath}
\usepackage{amssymb}
\usepackage{epsfig}
\usepackage{color}
\usepackage{graphicx}
\usepackage{hyperref}
\allowdisplaybreaks

\begin{document}

\title{Contributions of $K_0^*(1430)$ and $K_0^*(1950)$ in the charmed three-body $B$ meson decays}
\author{Bo-Yan Cui$^1$}
\author{Ya-Hui Chen$^2$}\email{yahuierdongchen@163.com}

\affiliation{$^1$School of Physics, Nankai University, Weijin Road 94, 300071 Tianjin, P.R. China}
\affiliation{$^2$Department of Physics, College of Basic Medical Sciences, Army Medical University, Chongqing 400038, P.R. China}

\date{\today}

\begin{abstract}
In this work, we investigate the resonant contributions of $K_0^*(1430)$ and $K_0^*(1950)$ in the three-body $B_{(s)}\to D_{(s)}K\pi$ within the perturbative QCD approach. The form factor  $F_{k\pi}(s)$ are adopted to describe
the nonperturbative dynamics of the S-wave $K\pi$ system. The branching ratios of all concerned decays
are calculated and predicted to be in the order of $10^{-10}$ to $10^{-5}$. The ratio $R$ of branching fractions
between $B^0\to \bar{D}^0 K_0^{*0}(1430) \to \bar{D}^0K^+\pi^-$ and
$B_s^0 \to \bar{D}^0 \bar{K}_0^{*0}(1430)\to \bar{D}^0K^-\pi^+$ are predicted to be 0.0552,
which implies the discrepancy for the LHCb measurements. We expect that the predictions in this
work can be tested by the future experiments, especially, to resolve $R$ ratio discrepancy.
\end{abstract}

\pacs{13.20.He, 13.25.Hw, 13.30.Eg}
\maketitle

\section{INTRODUCTION}
Decays of the type $B \to D hh^\prime$, where a $B$ meson decays to a charmed meson and two light pseudoscalar mesons,
have attracted people's attention in recent years. On the one hand, the studies of these three-body processes have shown the
 potential to constrain the parameters of the unitarity triangle. For instance, the decay $B^0 \to \bar{D}^0\pi^+\pi^-$ is
sensitive to measure the CKM angle $\beta$~\cite{Charles:1998vf,Latham:2008zs,LHCb:2015klp}, while Dalitz plot analysis of the decays $B^0 \to \bar{D^0}K^+\pi^-$
and $B_s^0 \to \bar{D}^0K^+K^-$can further improve the determination of the CKM angle $\gamma$~\cite{Gershon:2009qc,LHCb:2015tsv,Nandi:2011uw,LHCb:2018oeb}.
On the other hand, the $B \to D hh^\prime$ decays provide opportunities for probing the rich resonant structure in the final states,
including the spectroscopy of charmed mesons and the components in two light mesons system. A series of results in this area have been acquired from the measurements performed by the Belle~\cite{Belle:2002cli,Belle:2006wbx,Belle:2014agw}, BaBar~\cite{BaBar:2004kuq,BaBar:2007qow,BaBar:2007xlt,BaBar:2009pnd}
and LHCb~\cite{LHCb:2015klp,LHCb:2015tsv,LHCb:2018oeb,LHCb:2012fze,LHCb:2014ioa,LHCb:2015eqv,LHCb:2015jfh,LHCb:2016lxy,LHCb:2017vtv} Collaborations.

In theory, a direct analysis of the three-body $B$ decays is particularly difficult on account of the entangled resonant and nonresonant contributions,
the complex interplay between the weak processes and the low-energy strong interactions~\cite{Charles:2017ptc}, and other possible final state interactions~\cite{Bediaga:2015mia,Bediaga:2017axw}.
Fortunately, most of three-body hadronic $B$ meson decay processes are considered to be dominated by the low-energy $S$-, $P$- and $D$-wave resonant states, which could be treated in the quasi-two-body
framework. By neglecting the interactions between the meson pair originated from the resonant states and the bachelor particle in the final states, the factorization theorem
is still valid as in the two-body case~\cite{Amato:2016xjv,Boito:2017jav}, and substantial theoretical efforts for different quasi-two-body $B$ meson decays has been made within
different theoretical approaches~\cite{Gronau:2003ep,Engelhard:2005hu,Gronau:2005ax,Imbeault:2011jz,
Gronau:2013mda,Bhattacharya:2013cvn,Bhattacharya:2014eca,Xu:2013rua,Xu:2013dta,
He:2014xha,Furman:2005xp,El-Bennich:2006rcn,El-Bennich:2009gqk,Leitner:2010ai,
Dedonder:2010fg,Cheng:2005ug,Cheng:2007si,Cheng:2013dua,Cheng:2014uga,Li:2014oca,
Cheng:2016shb,Krankl:2015fha}. As well, the contributions from various intermediate resonant state for the three-body decays $B \to D hh^\prime$ have been investigated in
Ref.~\cite{Ma:2016csn,Ma:2017kec,Ma:2019qlm,Ma:2020dvr,Ma:2020jsb,Ma:2022hhp}.

The understanding of the scalar mesons is a difficult and long-standing issue~\cite{ParticleDataGroup:2020ssz}. The scalar resonances usually have large decay widths which make them overlap strongly with the background. In the specific regions, such as the $K \bar{K}$ and $\eta\eta$ thresholds, cusps in the line shapes
of the near-by resonances will appear due to the contraction of the phase space. Moreover, the inner natures of scalars are still not completely clear.
Part of them, especially the ones below $1$ GeV, have also been interpreted as glueballs, meson-meson bound states or multi-quark states, besides the traditional quark-antiquark
configurations~\cite{Jaffe:1976ih,Weinstein:1982gc,Weinstein:1983gd,Weinstein:1990gu,Alford:2000mm,Close:2002zu,Maiani:2004uc,Amsler:2004ps,Bugg:2004xu,Achasov:2005hm,Klempt:2007cp,Pelaez:2015qba}.
The $K_0^*(1430)$ is perhaps the least controversial of the light scalar mesons and generally believed to be a $q\bar{q}$ state~\cite{Anisovich:1997qp}.
It predominantly couples to the $K\pi$ channel and has been studied experimentally in many charmless three-body $B$ meson decays~\cite{Belle:2005rpz,BaBar:2007eog,Belle:2006ljg,BaBar:2008lpx,
BaBar:2009jov,BaBar:2015pwa,LHCb:2019xmb,LHCb:2019vww}. Recently, measurements of the charmed three-body decays $B^0 \to \bar{D}^0K^+\pi^-$ and
$B_s^0 \to \bar{D}^0K^+\pi^-$  involving the resonant state $K_0^*(1430)$ were also presented by LHCb~\cite{LHCb:2015tsv,LHCb:2014ioa}.
In addition, the subprocess $K_0^*(1950) \to K\pi$ which often ignored in literatures has also been considered in Ref.~\cite{LHCb:2014ioa}.

In the framework of the PQCD approach~\cite{Keum:2000ph,Keum:2000wi,Lu:2000em}, the investigation of $S$-wave $K\pi$ contributions to the $B^0_{(s)} \to \psi K\pi$ decays was carried out in Ref.~\cite{Rui:2017hks}.
In a more recent work~\cite{Wang:2020saq}, contributions of the resonant states $K_0^*(1430)$ and $K_0^*(1950)$ in the three-body decays  $B \to K\pi h$ ($h=K,\pi$) were studied systematically
within the same method. The $K_0^*(1430)$ is treated as the lowest lying $q\bar{q}$ state in view of the controversy for $K_0^*(700)$,
and the scalar $K\pi$ timelike form  factor $F_{K\pi}(s)$ was also discussed in detail. Motivated by the related results measured by LHCb\cite{LHCb:2015tsv,LHCb:2014ioa},
we shall extent the previous work~\cite{Wang:2020saq} to the study of the charmed three-body $B$ decays and analyse the contributions of the resonances $K_0^*(1430)$ and $K_0^*(1950)$ in
the $B \to DK\pi$ decays in this work.

The rest of this article is structured as follows. In Sec.~\ref{sec:2}, we give a brief review of the framework of the PQCD approach.
The numerical results and phenomenological discussions are presented in Sec.~\ref{sec:3} and a short summary is given in Sec.~\ref{sec:4}, respectively.
Finally, the relevant factorization formulae for the decay amplitudes are collected in the Appendix.

\section{FRAMEWORK}\label{sec:2}
\begin{figure}[tbp]
\centerline{\epsfxsize=13cm \epsffile{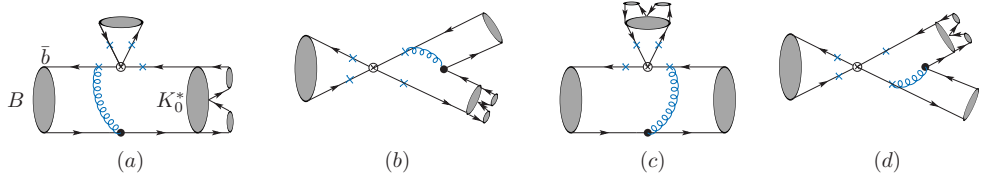}}
\vspace{0.3cm}
\caption{Typical diagrams for the quasi-two-body decays $B_{(s)}\to D_{(s)}K^*_0(1430,1950)\to D_{(s)}K\pi$ including the emission diagram (a) with the $B\to K^*_0(1430,1950)$ transition,
 the emission diagram (c) with the $B\to D$ transition, and the annihilation diagrams (b) and (d). The symbol $\otimes$ stands
for the weak vertex and $\times$ denotes possible attachments of hard gluons.}
\label{fig-FD}
\end{figure}

In the light-cone coordinate system, the $B$ meson momentum $p_B$, the total momentum of the $K\pi$ pair and the $D$ meson momentum $p_3$ under the rest frame of $B$ meson can be written as
\begin{eqnarray}
p_B=\frac{m_B}{\sqrt2}(1,1,\textbf{0}_{\rm T}),~~~~~~~~~~
p=\frac{m_B}{\sqrt2}(1-r^2,\eta,\textbf{0}_{\rm T}),~~~~~~~~~~
p_3=\frac{m_B}{\sqrt2}(r^2,1-\eta,\textbf{0}_{\rm T}),
\end{eqnarray}
with $m_B$ being the $B$ meson mass and the mass ratio $r=m_D/m_B$. The variable $\eta$ equals to $s/(m_B^2-m_D^2)$ where $s$ is the
invariant mass squared of $K\pi$ pair in the range from $(m_K+m_{\pi})^2$ to $(m_B-m_D)^2$. We also set the momenta of the light quarks in the $B$ meson, the $K\pi$ pair and the $D$ meson
as $K_B$, $K$ and $K_3$, and have the definitions as follow
\begin{eqnarray}
k_B=\left(0,\frac{m_B}{\sqrt2}x_B ,\textbf{k}_{B{\rm T}}\right),~~~~
k=\left(\frac{m_B}{\sqrt2}(1-r^2)z,0,\textbf{k}_{\rm T}\right),~~~~
k_3=\left(0,\frac{m_B}{\sqrt2}(1-\eta)x_3,\textbf{k}_{3{\rm T}}\right),
\end{eqnarray}
where $x_B$, $z$ and $x_3$ are the momentum fractions and run from zero to unity.

In the PQCD approach, the decay amplitude for the quasi-two-body decay $B_{(s)}\to D_{(s)}K^*_0(1430,1950)\to D_{(s)}K\pi$ can be expressed as the convolution~\cite{Chen:2002th}
\begin{eqnarray}
\mathcal{A}=\phi_B \otimes H \otimes \phi_{D} \otimes \phi_{K\pi},
\end{eqnarray}
where the symbol $H$ represents the hard kernel with single hard gluon exchange. $\phi_B$ and $\phi_{D}$ are the distribution amplitudes for the $B$ and $D$ mesons, respectively.
$\phi_{K\pi}$ denotes the distribution amplitude for the $K\pi$ pair with certain spin in the resonant region. In this work, we use the same distribution amplitudes for the $B_{(s)}$ and $D_{(s)}$
mesons as in Ref~\cite{Ma:2019qlm} where one can easily find their expressions and the relevant parameters.
Inspired by generalized distribution amplitude~\cite{Muller:1994ses,Diehl:1998dk,Polyakov:1998ze,Hagler:2002nh},
the generalized LCDA for two-meson system are introduced~\cite{Chen:2002th,Meissner:2013hya}
for three-body $B$-meson decay in the framework of PQCD approach and the heavy-to-light transition form factor in light-cone sum rules, respectively. The nonlocal matrix elements of vacuum to $K\pi$ with various spin projector can be written as
\begin{eqnarray}
\langle K \pi |\bar{s}(x) \, \gamma_{\mu} \,q(0) |0\rangle &=&p_{\mu} \int^1_0 \, dz \, e^{i\,z\, p\cdot x}\,  \phi^0(z,s)\,,\\
\langle K \pi |\bar{s}(x)  \,q(0) |0\rangle &=&\sqrt{s} \int^1_0 \, dz \, e^{i\,z\, p\cdot x}\,  \phi^s(z,s)\,,\\
\langle K \pi |\bar{s}(x) \, \sigma_{\mu\nu} \,q(0) |0\rangle &=&-\frac{\sqrt{s}}{6}\,(p_{\mu}x_{\nu}-p_{\nu}x_{\mu}) \int^1_0 \, dz \, e^{i\,z\, p\cdot x}\,  \phi^t(z,s)\, ,
\end{eqnarray}
the $K\pi$ $S$-wave distribution amplitude is chosen as~\cite{Wang:2020saq}
\begin{eqnarray}
\Phi_{K\pi}(z,s)=\frac{1}{\sqrt{2N_c}}\bigg[p\hspace{-1.6truemm}/ \phi^0(z,s)+
\sqrt{s} \phi^{s}(z,s)+\sqrt{s}(n\hspace{-2.0truemm}/ v\hspace{-1.8truemm}/ -1) \phi^t(z,s) \bigg],
\end{eqnarray}
where $n=(1,0,\mathbf{0}_T)$ and $v=(0,1,\mathbf{0}_T)$ are the dimensionless lightlike unit vectors.
The twist-$2$ and twist-$3$ light-cone distribution amplitudes have the form
\begin{eqnarray}
\phi^0(z,s)&=&\frac{F_{K\pi}(s)}{2\sqrt{2N_c}}6z(1-z)\bigg[a_0(\mu)
+\sum_{m=1}^{\infty}a_m(\mu) C_m^{3/2} (2z-1)\bigg],\nonumber\\
\phi^s(z,s)&=&\frac{F_{K\pi}(s)}{2\sqrt{2N_c}}, \nonumber\\
\phi^t(z,s)&=&\frac{F_{K\pi}(s)}{2\sqrt{2N_c}}(1-2z).
\end{eqnarray}
Here, $C_m^{3/2}$ are the Gegenbauer polynomials, $a_m(\mu)$ are the Gegenbauer moments and $F_{K\pi}(s)$ is the scalar form factor for the $K\pi$ pair.
In this work, we adopt the same formulae and parameters for the $K\pi$ $S$-wave distribution amplitude as them in Ref.~\cite{Wang:2020saq}.

According to the typical Feynman diagrams as shown in Fig.~\ref{fig-FD} and the quark currents for each decays,
the decay amplitudes for the considered quasi-two-body decays $B\to DK_{0}^{*}(1430,1950)\to DK\pi$ are given as
\begin{eqnarray}
{\mathcal A}\big(B^+\to D^0[K_0^{*+}\to]K\pi\big)&=&\frac{G_F}{\sqrt2}V^*_{ub}V_{cs}
\bigg\{a_2 F_{TK}+C_2 M_{TK}+a_1 F_{AD}+C_1 M_{AD}\bigg\}
 \;,\\
{\mathcal A}\big(B^+\to \bar{D}^0[K_0^{*+}\to]K\pi\big)&=&\frac{G_F}{\sqrt2}V^*_{cb}V_{us}
\bigg\{a_2 F_{TK}+C_2 M^{\prime}_{TK}+a_1 F_{TD}+C_1 M_{TD}\bigg\}
 \;,\\
{\mathcal A}\big(B^+\to D^+[K_0^{*0}\to]K\pi\big)&=&\frac{G_F}{\sqrt2}V^*_{ub}V_{cs}
\bigg\{a_1 F_{AD}+C_1 M_{AD}\bigg\}
 \;,\\
{\mathcal A}\big(B^+\to D_s^+[\bar{K}_0^{*0}\to]K\pi\big)&=&\frac{G_F}{\sqrt2}V^*_{ub}V_{cd}
\bigg\{a_1 F_{AD}+C_1 M_{AD}\bigg\}
 \;,\\
{\mathcal A}\big(B^0\to D^0[K_0^{*0}\to]K\pi\big)&=&\frac{G_F}{\sqrt2}V^*_{ub}V_{cs}
\bigg\{a_2 F_{TK}+C_2 M_{TK}\bigg\}
 \;,\\
{\mathcal A}\big(B^0\to \bar{D}^0[K_0^{*0}\to]K\pi\big)&=&\frac{G_F}{\sqrt2}V^*_{cb}V_{us}
\bigg\{a_2 F_{TK}+C_2 M^{\prime}_{TK}\bigg\}
 \;,\\
{\mathcal A}\big(B^0\to D^-[K_0^{*+}\to]K\pi\big)&=&\frac{G_F}{\sqrt2}V^*_{cb}V_{us}
\bigg\{a_1 F_{TD}+C_1 M_{TD}\bigg\}
 \;,\\
{\mathcal A}\big(B^0\to D_s^-[K_0^{*+}\to]K\pi\big)&=&\frac{G_F}{\sqrt2}V^*_{cb}V_{ud}
\bigg\{a_2 F_{AK}+C_2 M_{AK}\bigg\}
 \;,\\
{\mathcal A}\big(B^0\to D_s^+[K_0^{*-}\to]K\pi\big)&=&\frac{G_F}{\sqrt2}V^*_{ub}V_{cd}
\bigg\{a_2 F_{AD}+C_2 M_{AD}\bigg\}
 \;,\\
{\mathcal A}\big(B_s\to D^0[\bar{K}_0^{*0}\to]K\pi\big)&=&\frac{G_F}{\sqrt2}V^*_{ub}V_{cd}
\bigg\{a_2 F_{TK}+C_2 M_{TK}\bigg\}
 \;,\\
{\mathcal A}\big(B_s\to \bar{D}^0[\bar{K}_0^{*0}\to]K\pi\big)&=&\frac{G_F}{\sqrt2}V^*_{cb}V_{ud}
\bigg\{a_2 F_{TK}+C_2 M^{\prime}_{TK}\bigg\}
 \;,\\
{\mathcal A}\big(B_s\to D^+[K_0^{*-}\to]K\pi\big)&=&\frac{G_F}{\sqrt2}V^*_{ub}V_{cd}
\bigg\{a_1 F_{TK}+C_1 M_{TK}\bigg\}
 \;,\\
{\mathcal A}\big(B_s\to D_s^-[K_0^{*+}\to]K\pi\big)&=&\frac{G_F}{\sqrt2}V^*_{cb}V_{us}
\bigg\{a_1 F_{TD}+C_1 M_{TD}+a_2 F_{AK}+C_2 M_{AK}\bigg\}
 \;,\\
{\mathcal A}\big(B_s\to D_s^+[K_0^{*-}\to]K\pi\big)&=&\frac{G_F}{\sqrt2}V^*_{ub}V_{cs}
\bigg\{a_1 F_{TK}+C_1 M_{TK}+a_2 F_{AD}+C_2 M_{AD}\bigg\}
 \;,
\end{eqnarray}
where $G_F$ is the Fermi constant, $V_{ij}$ is the CKM matrix element, and the
combinations of the Wilson coefficients $a_{1,2}$ are defined as $a_1=C_1/3+C_2$ and $a_2=C_2/{3}+C_1$.
The expressions of individual amplitudes $F_{TK}$, $F_{TD}$, $F_{AK}$, $F_{AD}$, $M_{TK}^{(\prime)}$, $M_{TD}$, $M_{AK}$ and $M_{AD}$
from different subdiagrams in Fig.~\ref{fig-FD} are collected in the Appendix.

At last, we give the definition of the differential branching ratio for the considered quasi-two-body decays
\begin{equation}\label{DBR}
\frac{d\mathcal{B}}{ds}=\tau_B\, \frac{|\vec{p}_1||\vec{p}_3|}{64\pi^3m_B^3}\, |\mathcal{A}|^2.
\end{equation}
In the center-of-mass frame of $K\pi$ system,  the magnitudes of the momenta $|\vec{p}_1|$ and $|\vec{p}_3|$ can be expressed as
\begin{eqnarray}
|\vec{p}_1|&=&\frac{1}{2} \sqrt{[(m^2_{K}-m^2_\pi)^2-2(m^2_{K}+m^2_\pi)s+s^2]/s}, \nonumber\\
|\vec{p}_3|&=&\frac{1}{2} \sqrt{[(m^2_{B}-m^2_D)^2-2(m^2_{B}+m^2_D)s+s^2]/s}.
\end{eqnarray}

\section{RESULTS}\label{sec:3}
In the numerical calculations, the masses of the involved mesons (GeV), the lifetime of the $B$ mesons (ps), the resonance decay widths (GeV)
and the Wolfenstein parameters
are taken from the $Review~of~Particle~Physics$~\cite{ParticleDataGroup:2020ssz}
\begin{eqnarray}\label{PDG}
m_{B^\pm}&=&5.279,\quad m_{B^0}=5.280,\quad m_{B_{s}^{0}}=5.367,\quad m_{D^0/\bar{D}^0}=1.865, \quad m_{D^{\pm}}=1.870,\quad m_{D_s^{\pm}}=1.968,\nonumber\\
m_{K^{\pm}}&=&0.494, \quad m_{K^0/\bar{K}^0}=0.498, \quad m_{\pi^0}=0.135, \quad m_{\pi^{\pm}}=0.140,\quad  m_{K_0^*(1430)}=1.425,\quad m_{K_0^*(1950)}=1.945,\nonumber\\
\tau_{B^{0}}&=&1.519,\quad \tau_{B^{\pm}}=1.638,\quad \tau_{B_{s}^{0}}=1.515, \quad \Gamma_{K_0^*(1430)}=0.270\pm 0.080,\quad \Gamma_{K_0^*(1950)}=0.201\pm 0.090,\nonumber\\
 A&=&0.790^{+0.017}_{-0.012},\quad \lambda=0.22650\pm0.00048,\quad \bar{\rho}=0.141^{+0.016}_{-0.017}\quad \bar{\eta}=0.357\pm0.011.
\end{eqnarray}
The decay constants of the $B_{(s)}$ and $D_{(s)}$ mesons are set to the values $f_{B_{(s)}}=0.190~(0.230)$ GeV and
$f_{D_{(s)}}=0.212~(0.250)$ GeV~\cite{Aoki:2021kgd}.

By integrating the differential branching ratio in Eq.~(\ref{DBR}), we obtain the branching ratios for the considered quasi-two-body processes with the intermediate
resonances $K_0^*(1430)$ and $K_0^*(1950)$ in Tables~\ref{B2DK1430} and \ref{B2DK1950}, respectively.
The first error is induced by the shape parameters $\omega_{B_{(s)}}=0.40 \pm 0.04~(0.50 \pm 0.05)$ GeV in the distribution amplitude for the $B_{(s)}$ meson.
The second and third errors come from the Gegenbauer moments $a_3=-0.42\pm 0.22$ and $a_1=-0.57 \pm 0.13$ in the $K\pi$ $S$-wave distribution amplitude, respectively.
The decay widths $\Gamma_{K^*_0(1430)}=0.270\pm 0.080$ GeV and $\Gamma_{K^*_0(1950)}=0.201\pm 0.090$ GeV contribute the fourth error.
The last one is due to the parameter $C_{D_{(s)}}=0.5\pm0.1~(0.4\pm0.1)$ in the distribution amplitude for $D_{(s)}$ meson.
The uncertainties from other parameters are comparatively small and have been neglected.

\begin{table}[thb]
\begin{center}
\caption{PQCD predictions for the branching fractions of the quasi-two-body decays $B\to D_{(s)}K_0(1430)^*\to D_{(s)}K\pi$ together with the available experimental data.}
\begin{tabular}{l c c l}
  \hline\hline
  \     ~~~~~~~~Mode       & Unit & $\mathcal{B}$  &  ~~~~Data  \\  \hline
  $B^+\to D^0 K_0^{*+}(1430) \to D^0K^0\pi^+$           ~&~ ($10^{-6}$) ~&~ $4.15 \pm 0.30(\omega_B) \pm 0.16(B_3) \pm 0.15(B_1) \pm 0.25 (\Gamma_{K_0^*}) \pm 0.02 (C_D)$ ~&~  -  \\

  $B^+\to \bar{D}^0 K_0^{*+}(1430) \to \bar{D}^0K^0\pi^+$           ~&~ ($10^{-5}$) ~&~ $2.50 \pm 0.17(\omega_B) \pm 0.28(B_3)\pm 0.04(B_1) \pm 0.20 (\Gamma_{K_0^*}) \pm 0.03 (C_D)$ ~&~  -  \\

  $B^+\to D^+ K_0^{*0}(1430) \to D^+K^+\pi^-$     ~&~ ($10^{-8}$) ~&~ $2.14 \pm 0.87(\omega_B) \pm 0.55(B_3)\pm 0.30(B_1) \pm 0.07 (\Gamma_{K_0^*}) \pm 0.05 (C_D)$ ~&~  - \\

  $B^+\to D_s^+ \bar{K}_0^{*0}(1430) \to D_s^+K^-\pi^+$  ~&~ ($10^{-9}$) ~&~ $2.75 \pm 0.70(\omega_B) \pm 1.18(B_3)\pm 0.68(B_1) \pm 0.06 (\Gamma_{K_0^*}) \pm 0.22 (C_D)$ ~&~  -   \\

  $B^0\to D^0 K_0^{*0}(1430) \to D^0K^+ \pi^-$           ~&~ ($10^{-6}$) ~&~ $3.90 \pm 0.28(\omega_B) \pm 0.01(B_3) \pm 0.13(B_1) \pm 0.23 (\Gamma_{K_0^*}) \pm 0.01 (C_D)$ ~&~ -  \\

  $B^0\to \bar{D}^0 K_0^{*0}(1430) \to \bar{D}^0K^+\pi^-$  ~&~ ($10^{-5}$) ~&~ $2.23 \pm 0.18(\omega_B) \pm 0.24(B_3) \pm 0.06(B_1) \pm 0.15 (\Gamma_{K_0^*}) \pm 0.05 (C_D)$ ~&~  0.71~\cite{LHCb:2015tsv}  \\

  $B^0\to D^- K_0^{*+}(1430)\to D^-K^0\pi^+$             ~&~ ($10^{-7}$) ~&~ $1.08 \pm 0.17(\omega_B) \pm 0.34(B_3) \pm 0.12(B_1)\pm 0.06 (\Gamma_{K_0^*}) \pm 0.02 (C_D)$ ~&~ -  \\

  $B^0\to D_s^- K_0^{*+}(1430)\to D_s^-K^0\pi^+$ ~&~ ($10^{-6}$) ~&~ $2.17 \pm 1.08(\omega_B) \pm 1.20(B_3)\pm 0.75(B_1) \pm 0.10 (\Gamma_{K_0^*}) \pm 0.10 (C_D)$ ~&~ -  \\

  $B^0\to D_s^+ K_0^{*-}(1430)\to D_s^+\bar{K}^0\pi^-$ ~&~ ($10^{-9}$) ~&~ $4.24 \pm 1.99(\omega_B) \pm 1.97(B_3) \pm 0.61(B_1)\pm 0.14 (\Gamma_{K_0^*}) \pm 0.09 (C_D)$ ~&~ -  \\

  $B_s^0\to D^0 \bar{K}_0^{*0}(1430)\to D^0K^-\pi^+$ ~&~ ($10^{-7}$) ~&~ $2.07 \pm 0.17(\omega_B) \pm 0.20(B_3) \pm 0.11(B_1)\pm 0.13 (\Gamma_{K_0^*}) \pm 0.01 (C_D)$ ~&~ -  \\

  $B_s^0\to \bar{D}^0 \bar{K}_0^{*0}(1430)\to \bar{D}^0K^-\pi^+$ ~&~ ($10^{-4}$) ~&~ $3.76 \pm 0.16(\omega_B) \pm 0.43(B_3)\pm 0.08(B_1) \pm 0.23 (\Gamma_{K_0^*}) \pm 0.03 (C_D)$ ~&~ 3.00~\cite{LHCb:2014ioa}  \\

  $B_s^0\to D^+ K_0^{*-}(1430)\to D^+ \bar{K}^0\pi^-$ ~&~ ($10^{-6}$) ~&~ $7.67 \pm 1.71(\omega_B) \pm 0.43(B_3)\pm 0.32(B_1) \pm 0.48 (\Gamma_{K_0^*}) \pm 0.01 (C_D)$ ~&~ -  \\

  $B_s^0\to D_s^- K_0^{*+}(1430)\to D_s^-K^0\pi^+$ ~&~ ($10^{-7}$) ~&~ $1.65 \pm 0.28(\omega_B) \pm 0.25(B_3)\pm 0.16(B_1) \pm 0.26 (\Gamma_{K_0^*}) \pm 0.26 (C_D)$ ~&~ -  \\

  $B_s^0\to D_s^+ K_0^{*-}(1430)\to D_s^+\bar{K}^0\pi^-$ ~&~ ($10^{-4}$) ~&~ $1.96 \pm 0.44(\omega_B) \pm 0.16(B_3) \pm 0.08(B_1) \pm 0.12 (\Gamma_{K_0^*}) \pm 0.01 (C_D)$ ~&~ -  \\
  \hline\hline
  \label{B2DK1430}
\end{tabular}
\vspace{-0.5cm}
\end{center}
\end{table}

\begin{table}[thb]
\begin{center}
\caption{PQCD predictions for the branching fractions of the quasi-two-body decays $B\to D_{(s)}K_0(1950)^*\to DK\pi$ together with the available experimental data.}
\begin{tabular}{l c c l}
  \hline\hline
  \     ~~~~~~~~Mode       & Unit & $\mathcal{B}$  &   ~~~~Data  \\  \hline
  $B^+\to D^0 K_0^{*+}(1950) \to D^0K^0\pi^+$           ~&~ ($10^{-7}$) ~&~ $1.39 \pm 0.60(\omega_B) \pm 0.13(B_3) \pm 0.12(B_1)\pm 0.04 (\Gamma_{K_0^*}) \pm 0.04 (C_D)$ ~&~  -  \\

  $B^+\to \bar{D}^0 K_0^{*+}(1950) \to \bar{D}^0K^0\pi^+$           ~&~ ($10^{-7}$) ~&~ $5.52 \pm 2.70(\omega_B) \pm 0.57(B_3) \pm 0.08(B_1)\pm 0.34 (\Gamma_{K_0^*}) \pm 0.12 (C_D)$ ~&~  -  \\

  $B^+\to D^+ K_0^{*0}(1950) \to D^+K^+\pi^-$     ~&~ ($10^{-9}$) ~&~ $1.31 \pm 0.37(\omega_B) \pm 0.17(B_3) \pm 0.10(B_1) \pm 0.10 (\Gamma_{K_0^*}) \pm 0.12 (C_D)$ ~&~  - \\

  $B^+\to D_s^+ \bar{K}_0^{*0}(1950) \to D_s^+K^-\pi^+$    ~&~ ($10^{-10}$) ~&~ $1.94 \pm 0.44(\omega_B) \pm 0.82(B_3) \pm 0.44(B_1) \pm 0.02 (\Gamma_{K_0^*}) \pm 0.21 (C_D)$ ~&~  -   \\

  $B^0\to D^0 K_0^{*0}(1950) \to D^0K^+ \pi^-$           ~&~ ($10^{-7}$) ~&~ $1.30 \pm 0.57(\omega_B) \pm 0.16(B_3) \pm 0.12(B_1) \pm 0.07 (\Gamma_{K_0^*}) \pm 0.01 (C_D)$ ~&~ -  \\

  $B^0\to \bar{D}^0 K_0^{*0}(1950) \to \bar{D}^0K^+\pi^-$   ~&~ ($10^{-7}$) ~&~ $5.23 \pm 2.14(\omega_B) \pm 0.23(B_3) \pm 0.06(B_1) \pm 0.24 (\Gamma_{K_0^*}) \pm 0.14 (C_D)$ ~&~ -  \\

  $B^0\to D^- K_0^{*+}(1950)\to D^-K^0\pi^+$             ~&~ ($10^{-9}$) ~&~ $4.32 \pm 0.38(\omega_B) \pm 1.21(B_3)\pm 0.42(B_1) \pm 0.24 (\Gamma_{K_0^*}) \pm 0.43 (C_D)$ ~&~ -  \\

  $B^0\to D_s^- K_0^{*+}(1950)\to D_s^-K^0\pi^+$ ~&~ ($10^{-7}$) ~&~ $1.04 \pm 0.57(\omega_B) \pm 0.70(B_3) \pm 0.29(B_1) \pm 0.03 (\Gamma_{K_0^*}) \pm 0.05 (C_D)$ ~&~ -  \\

  $B^0\to D_s^+ K_0^{*-}(1950)\to D_s^+\bar{K}^0\pi^-$ ~&~ ($10^{-10}$) ~&~ $2.59 \pm 1.18(\omega_B) \pm 1.05(B_3) \pm 0.49(B_1)\pm 0.10 (\Gamma_{K_0^*}) \pm 0.07 (C_D)$ ~&~ -  \\

  $B_s^0\to D^0 \bar{K}_0^{*0}(1950)\to D^0K^-\pi^+$ ~&~ ($10^{-9}$) ~&~ $9.77 \pm 2.68(\omega_B) \pm 0.69(B_3) \pm 0.42(B_1)\pm 0.38 (\Gamma_{K_0^*}) \pm 0.08 (C_D)$ ~&~ -  \\

  $B_s^0\to \bar{D}^0 \bar{K}_0^{*0}(1950)\to \bar{D}^0K^-\pi^+$ ~&~ ($10^{-5}$) ~&~ $1.72 \pm 0.44(\omega_B) \pm 0.18(B_3) \pm 0.04(B_1) \pm 0.05 (\Gamma_{K_0^*}) \pm 0.01 (C_D)$ ~&~ $<11$~\cite{LHCb:2014ioa}  \\

  $B_s^0\to D^+ K_0^{*-}(1950)\to D^+ \bar{K}^0\pi^-$ ~&~ ($10^{-7}$) ~&~ $3.48 \pm 1.50(\omega_B) \pm 0.10(B_3)\pm 0.07(B_1) \pm 0.16 (\Gamma_{K_0^*}) \pm 0.01 (C_D)$ ~&~ -  \\

  $B_s^0\to D_s^- K_0^{*+}(1950)\to D_s^-K^0\pi^+$ ~&~ ($10^{-9}$) ~&~ $1.12 \pm 0.28(\omega_B) \pm 0.11(B_3) \pm 0.07(B_1) \pm 0.8 (\Gamma_{K_0^*}) \pm 0.03 (C_D)$ ~&~ -  \\

  $B_s^0\to D_s^+ K_0^{*-}(1950)\to D_s^+\bar{K}^0\pi^-$ ~&~ ($10^{-6}$) ~&~ $8.37 \pm 3.71(\omega_B) \pm 0.22(B_3)\pm 0.12(B_1) \pm 0.39 (\Gamma_{K_0^*}) \pm 0.02 (C_D)$ ~&~ -  \\
  \hline\hline
  \label{B2DK1950}
\end{tabular}
\vspace{-0.5cm}
\end{center}
\end{table}

From the numerical results as listed in Tables~\ref{B2DK1430} and~\ref{B2DK1950}, we have the following comments:
\begin{enumerate}
  \item[(1)]
  In the $B\to DR\to DK\pi$ decays, we can extract the two-body branching fractions $\mathcal{B}(B\to DR)$
  by using the relation under the quasi-two-body approximation
  \begin{equation}\label{NWA}
  \mathcal{B}(B\to DR\to DK\pi)=\mathcal{B}(B\to DR) \cdot \mathcal{B}(R\to K\pi)\;.
  \end{equation}
  For the branching fractions of two-body decays with $K_0^*(1430)$ and $K_0^*(1950)$, we shall apply
  \begin{equation}
  \mathcal{B}(K_0^{*0}\to K^+\pi^-)=\mathcal{B}(\bar{K}_0^{*0}\to K^-\pi^+)=\mathcal{B}(K_0^{*+}\to K^0\pi^+)=
  \mathcal{B}(K_0^{*-}\to \bar{K}^0\pi^-) = \frac{2}{3}\mathcal{B}(K_0^*\to K\pi).
    \end{equation}
    and the values
  \begin{equation}
  \mathcal{B}(K_0^*(1430)\to K\pi)=(93 \pm 10)\%,~~~~~~\mathcal{B}(\bar{K}_0^*(1950)\to K^-\pi^+)=(52 \pm 14)\%.
  \label{BrHad}
  \end{equation}
  Combined with results listed in Tables~\ref{B2DK1430} and~\ref{B2DK1950}, one can obtain the related two-body branching fractions,
   for examples, $B_s^0\to \bar{D}^0 \bar{K}_0^{*0}(1430)=6.06\pm0.65\times 10^{-4}$ and $B_s^0\to \bar{D}^0 \bar{K}_0^{*0}(1950)=3.31\pm 0.89 \times 10^{-5}$, where the
   errors are propagated from eq.~(\ref{BrHad}).

  \item[(2)]
  The PQCD prediction for the branching fraction $\mathcal{B}(B_s^0\to \bar{D}^0 \bar{K}_0^{*0}(1430)\to \bar{D}^0K^-\pi^+)$ agrees with LHCb's data
    $(3.00\pm0.24\pm0.11\pm0.50\pm0.44)\times10^{-4}$~\cite{LHCb:2014ioa} 
    within errors, while the PQCD predicted
    $ \mathcal{B}(B^0\to \bar{D}^0 K_0^{*0}(1430) \to \bar{D}^0K^+\pi^-)$ is much larger than
    the value $(0.71\pm0.27\pm0.33\pm0.47\pm0.08)\times 10^{-5}$ measured by LHCb~\cite{LHCb:2015tsv}
    with significant uncertainties. By comparison,
    one can find that the decay modes
   $B_s^0\to \bar{D}^0 \bar{K}_0^{*0}(1430)\to \bar{D}^0K^-\pi^+$ and $B^0\to \bar{D}^0 K_0^{*0}(1430) \to \bar{D}^0K^+\pi^-$ contain the same decay topology
   when neglecting the differences of hadronic parameters between $B^0$ and $B_s^0$. Then, we evaluate the ratio
       \begin{equation}\label{R}
       R=\frac{\mathcal{B}(B^0\to \bar{D}^0 K_0^{*0}(1430) \to \bar{D}^0K^+\pi^-)}{\mathcal{B}(B_s^0 \to \bar{D}^0 \bar{K}_0^{*0}(1430)\to \bar{D}^0K^-\pi^+)}\approx \left|\frac{V_{us}}{V_{ud}}\right|^2
       \cdot \frac{\tau_{B^0}}{\tau_{B_s^0}}=0.0534\;,
       \end{equation}
     which is close to the PQCD prediction $0.0593$ by using the results listed in Table~\ref{B2DK1430}, 
     but different from the value $0.0237$ acquired from the central values of the measured
     branching ratio by LHCb~\cite{LHCb:2015tsv,LHCb:2014ioa}.
     One can find that in the Ref~\cite{LHCb:2014ioa}, 
     the $K_0^{*}(1430)$ component receive 20\% fit fraction of total $\mathcal{B}(B_s^0\to \bar{D}^0K^-\pi^+)$, but in Ref~\cite{LHCb:2015tsv,LHCb:2014ioa},
     $K_0^{*}(1430)$ component receive only 5.1\% of total $\mathcal{B}(B^0\to \bar{D}^0K^+\pi^-)$. The $K_0^{*}(1430)$ component playing a such different role in two different process, however, on the theoretical side, the decay amplitudes are exactly same for $B^0\to \bar{D}^0 K_0^{*0}(1430) \to \bar{D}^0K^+\pi^-$ and $B_s^0 \to \bar{D}^0 \bar{K}_0^{*0}(1430)\to \bar{D}^0K^-\pi^+$  if we neglect the SU(3) symmetry breaking effect, $R$ ratio will independent of theoretical framework. More precise measurements and more proper partial wave analysis are needed to resolve the discrepancy.

  \item[(3)]
For the CKM suppressed decay modes $B_s^0\to D^0 \bar{K}_0^{*0}(1430)\to D^0K^-\pi^+$, their branching ratios are
much smaller than the corresponding results of $B_s^0\to D^0 \bar{K}_0^{*0}(1430)\to D^0K^-\pi^+$ decays as predicted by
PQCD in this work. The major reason comes from the strong CKM suppression factor 
\begin{equation}
R_{CKM}=\bigg|\frac{V_{ub}^*V_{cd}}{V_{cb}^*V_{ud}}\bigg|^2\approx \lambda^4(\bar{\rho}^2+\bar{\eta}^2)\approx 3\times 10^{-4} \,,
\end{equation}
as discussed in Ref~\cite{Cui:2019khu}.
The non-vanishing charm quark mass in the fermion propagator generates the main differences between the $\frac{B_s^0\to D^0 \bar{K}_0^{*0}(1430)\to D^0K^-\pi^+}{B_s^0\to \bar{D}^0 \bar{K}_0^{*0}(1430)\to \bar{D}^0K^-\pi^+}$ and $R_{CKM}$. 
Similarly, for the $B^+\to D^0 K_0^{*+}(1430) \to D^0K^0\pi^+$ decay and $B^+\to \bar{D}^0 K_0^{*+}(1430) \to \bar{D}^0K^0\pi^+$ decay, there still exist the CKM suppression but much moderate  than the previous cases: 
\begin{equation}
    R^s_{CKM}=\bigg|\frac{V_{ub}^*V_{cs}}{V_{cb}^*V_{us}}\bigg|^2\approx(\bar{\rho}^2+\bar{\eta}^2)\approx 0.147
\end{equation}
from Table~\ref{B2DK1430} we have 
\begin{eqnarray}
R^{s1}_{CKM}=\frac{\mathcal{B}(B^+\to D^0 K_0^{*+}(1430) \to D^0K^0\pi^+)}{\mathcal{B}(B^+\to \bar{D}^0 K_0^{*+}(1430) \to \bar{D}^0K^0\pi^+)}&\approx& 0.166\,,
\\
R^{s2}_{CKM}=\frac{\mathcal{B}(B^0\to D^0 K_0^{*0}(1430) \to D^0K^+ \pi^-)}{\mathcal{B}(B^0\to \bar{D}^0 K_0^{*0}(1430) \to \bar{D}^0K^+\pi^-)}&\approx& 0.175\,.
\end{eqnarray}
The main differences between the $R^s_{CKM}$ and $R^{s1,s2}_{CKM}$ comes from the nonvanishing charm quark mass contributions in the non-factorizable $B\to K^*_0(1430)$ emission diagram. We also suggest more study on the decay mode $B_s^0\to D_s^+ K_0^{*-}(1430)\to D_s^+\bar{K}^0\pi^-$ because it has a  large branching ratio and can be found in  future experiments.

  \item[(4)] $K_0^*(1430)$ was often parameterized by LASS lineshape~\cite{Aston:1987ir} in partial wave analysis, which incorporate both cusps resonance and slowly varying nonresonance contribution, and it was applied in LHCb measurements~\cite{LHCb:2015tsv,LHCb:2014ioa}. However, rigorous theoretical calculation for nonresonance contribution in the context of PQCD framework is still absent~\cite{Wang:2020saq}, the comparison between theoretical calculations and experiment measurements focus only on the $S$-wave $K_0^*(1430)$ contribution. More attempts can be make in future study to parameterize the nonresonance contribution for sake of giving a more reliable result.
  
  \item[(5)]
    The $CP$-averaged branching fraction of the charmless quasi-two-body decay involving the intermediate state $K_0^*(1950)$ is predicted to be about one magnitude smaller than the corresponding process containing $K_0^*(1430)$ in~\cite{Wang:2020saq}. In quasi-two-body charmed decays, the ratio of branching fractions between Table~\ref{B2DK1950} and \ref{B2DK1430} are about few percentage, which are smaller than that of charmless cases mainly due to the absence of $(S-P)(S+P)$ amplitude, which receive resonance pole mass enhancement as discussed in~\cite{Wang:2020saq}. And the more compact phase space can also reduce the branching fractions for the decay mode involving $K_0^*(1950)$. From the partial wave analysis in~\cite{LHCb:2014ioa}, the $K_0^*(1950)$ mode is  measured to be about 1.5\% than that of $K_0^*(1430)$ mode , which is about one third of our prediction, ie. 4.6\%, more precise measurements and more reliable theoretical predictions are needed in the future study.

  \item[(6)] In Fig.~\ref{fig-wdep-D}, we show the $K\pi$ invariant mass-dependent differential branching fraction for the quasi-two-body decays $B_s^0\to\bar{D}_0^{*0}\bar{K}_0^{*0}(1430)\to \bar{D}_0^{*0}K^-\pi^+$ (solid line) and $B_s^0\to\bar{D}_0^{*0}\bar{K}_0^{*0}(1950)\to \bar{D}_0^{*0}K^-\pi^+$ (dashed line). One can easily find that the main portion of the branching fraction comes from the region around the pole mass of the corresponding resonant states, the contributions
      from the $m_{K\pi}$ mass region greater than 3 GeV is evaluated about 0.4\% compared with the whole kinematic region (i.e. $[m_K+m_{\pi},m_B-m_D]$)
      in this work and can be safely neglected.

\begin{figure}[tbp]
\centerline{\epsfxsize=8cm \epsffile{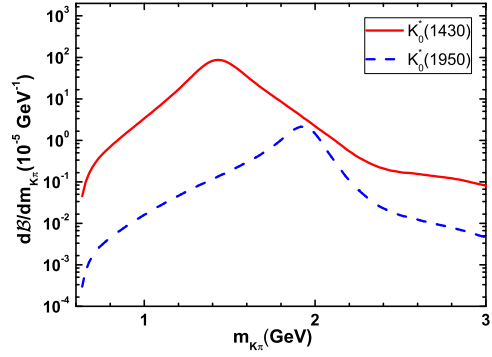}}
\vspace{0.3cm}
\caption{The $K\pi$ invariant mass-dependent differential branching fraction for $B_s^0\to\bar{D}_0^{*0}\bar{K}_0^{*0}(1430)\to \bar{D}_0^{*0}K^-\pi^+$ (solid line) and $B_s^0\to\bar{D}_0^{*0}\bar{K}_0^{*0}(1950)\to \bar{D}_0^{*0}K^-\pi^+$ (dashed line).}
\label{fig-wdep-D}
\end{figure}
\end{enumerate}

\section{CONCLUSION}\label{sec:4}
Motivated by the phenomenological importance of the charmed three-body hadronic $B$-meson decays, in the present work we have studied the quasi-two-body decays
$B_{(s)}\to D_{(s)}K^*_0(1430,1950)\to D_{(s)}K\pi$
in the PQCD factorization approach with the help of the scalar form factor
$F_{k\pi}(s)$ as a nonperturbative input. The branching ratios of all concerned decays are
calculated, and are of the order $10^{-10}$ to $10^{-5}$, the corresponding two-body branching fractions can be obtained by using the quasi-two-body approximation relation in Eq~(\ref{NWA}). Under SU(3) flavor symmetry, we found  the theoretical framework independent ratio
$R=\frac{\mathcal{B}(B^0\to \bar{D}^0 K_0^{*0}(1430) \to \bar{D}^0K^+\pi^-)}
{\mathcal{B}(B_s^0 \to \bar{D}^0 \bar{K}_0^{*0}(1430)\to \bar{D}^0K^-\pi^+)}
\approx \left|\frac{V_{us}}{V_{ud}}\right|^2
\cdot \frac{\tau_{B^0}}{\tau_{B_s^0}}\approx 0.0534$ by neglecting the differences
of hadronic parameters between $B^0$ and $B_s^0$, this result is consistent with our PQCD prediction,
but inconsistent with LHCb measurements. For the decays $B_s^0\to D^0 \bar{K}_0^{*0}(1430)\to D^0K^-\pi^+$ and $B_s^0\to D^0 \bar{K}_0^{*0}(1430)\to D^0K^-\pi^+$, the
great difference in their corresponding branching fractions can be understood by a strong CKM suppression factor $R_{CKM}\approx \lambda^4(\bar{\rho}^2+\bar{\eta}^2)\approx 3\times 10^{-4}$, while the moderate difference between 
$B^+\to D^0 K_0^{*+}(1430) \to D^0K^0\pi^+$ and $B^+\to \bar{D}^0 K_0^{*+}(1430) \to \bar{D}^0K^0\pi^+$ as well as 
$B^0\to D^0 K_0^{*0}(1430) \to D^0K^+ \pi^-$ and $B^0\to \bar{D}^0 K_0^{*0}(1430) \to \bar{D}^0K^+\pi^-$ are mainly due to the $R^s_{CKM}\approx(\bar{\rho}^2+\bar{\eta}^2)\approx 0.147$. More reliable theoretical predictions are
needed in the future study for the nonresonance contribution and $S$-wave $K_0^{*}(1950)$ contribution. We hope the predictions in this work can be tested by the future experiments, especially, to resolve $R$ ratio discrepancy.

\begin{acknowledgments}
We are grateful to Ai-jun Ma for helpful comments. This work was supported by the National Natural Science Foundation of China under Grant No. 11947040.
\end{acknowledgments}

\appendix*
\section{Decay Amplitudes}\label{appendix}
The factorization formulae for the individual amplitudes from different subdiagrams in Fig.~\ref{fig-FD} are
\begin{eqnarray}
F_{TK}&=&8\pi C_F m_B^4 f_D \int dx_B dz \int b_B db_B b db\phi_B(x_B,b_B)\big\{\big[\sqrt{\eta(1-r^2)}
[\phi^s(r^2(-2z\eta+2z+1)
\nonumber\\
&+&(\eta-1)(2z-1))-\phi^t\left(1+\eta+r^2(2(\eta-1)z+1)+2z(1-\eta)\right)]-\phi^0((\eta-1)r^4z
\nonumber\\
&+&r^2(-2\eta(z+1)+2z+1)+(\eta-1)(z+1)) \big]E_{1ab}(t_{1a})h_{1a}(x_B,z,b_B,b)S_t(z)
\nonumber\\
&+&\big[r^4\phi^0(\eta-x_B)+(\eta-1)(\eta\phi^0-2\phi^s\sqrt{\eta(1-r^2)})
+r^2[2\phi^s\sqrt{\eta(1-r^2)}(2\eta-1-x_B)
\nonumber\\
&+&(x_B-\eta^2)\phi^0] \big]E_{1ab}(t_{1b})h_{1b}(x_B,z,b_B,b)S_t(\vert x_B-\eta\vert) \big\},
\\
M_{TK}&=&32/\sqrt{6}\pi C_F m_B^4 \int dx_B dz dx_3 \int b_B db_B b_3 db_3 \phi_B(x_B,b_B) \phi_D \big\{\big[-\phi^0
(1+r^2-\eta)\nonumber\\
&\times&\left(\eta+r^2(\eta(2x_3+z-2)-x_3-x_B+1)-\eta(x_3+z)+x_3+x_B-1\right)
\nonumber\\
&+&\sqrt{\eta(1-r^2)}[r^2\left(\phi^s(2x_3+x_B+z-2-(2x_3+z-2)\eta)+\phi^t(x_B+z\eta-z)\right)
\nonumber\\
&+&(\eta-1)z(\phi^s-\phi^t)]\big]E_{1cd}(t_{1c})h_{1c}(x_B,z,x_3,b_B,b_3) +\big[z(2\eta-1)r^4\phi^0
\nonumber\\
&-&r^3r_c\phi^0+rr_c(\phi^0(1+\eta)-4\phi^s\sqrt{\eta(1-r^2)})+
(\eta-1)(z\sqrt{\eta(1-r^2)}(\phi^s+\phi^t)
\nonumber\\
&+&\phi^0(\eta x_3-x_3+x_B-z))+r^2[2(\eta-1)x_3(\eta\phi^0-\sqrt{\eta(1-r^2)}\phi^s)
\nonumber\\
&+&x_B\left((2\eta-1)\phi^0+\sqrt{\eta(1-r^2)}(\phi^t-\phi^s)\right)+
z\big((2-3\eta)\phi^0+(\eta-1)\sqrt{\eta(1-r^2)}
\nonumber\\
&\times& (\phi^t+\phi^s) \big)]\big] E_{1cd}(t_{1d})h_{1d}(x_B,z,x_3,b_B,b_3)\big\},
\\
M^{\prime}_{TK}&=&32/\sqrt{6}\pi C_F m_B^4 \int dx_B dz dx_3 \int b_B db_B b_3 db_3 \phi_B(x_B,b_B) \phi_D \big\{\big[
\sqrt{\eta(1-r^2)}\nonumber\\
&\times&[r^2\phi^s(x_3+z-2)(1-\eta)+r^2x_B\phi^s+r^2\phi^t(x_B+z\eta-z)
+4rr_c\phi^s+(\eta-1)z(\phi^s-\phi^t)]
\nonumber\\
&-&\phi^0\big(r^4(\eta(2x_3+z-2)+1-x_3-x_B)-r^3rc+\eta r^2(\eta(2-z-x_3)+x_3+x_B-2)
\nonumber\\
&+&(\eta+1)rr_c+(\eta-1)(\eta(x_3+z-1)-x_3-x_B+1) \big)\big]E_{1cd}(t^{\prime}_{1c})h^{\prime}_{1c}(x_B,z,x_3,b_B,b_3)
\nonumber\\
&+&\big[\phi^0(1+r^2(2\eta-1)-\eta)\big(x_B-x_3(1+r^2-\eta)+z(r^2-1) \big)+\sqrt{\eta(1-r^2)}
\nonumber\\
&\times& [\phi^s\big(2x_3r^2(1-\eta)-r^2x_B+(\eta-1)(r^2-1)z\big)+\phi^t(r^2x_B+(\eta-1)(r^2-1)z)]\big]
\nonumber\\
&\times&  E_{1cd}(t^{\prime}_{1d})h^{\prime}_{1d}(x_B,z,x_3,b_B,b_3)\big\},
\\
F_{AK}&=&8\pi C_F m_B^4 f_B \int dx_3 dz \int b_3 db_3 b db\phi_D\big\{\big[\sqrt{\eta(1-r^2)}[2r^3z(\phi^s-\phi^t)+ r^2r_c(\phi^t-\phi^s)
\nonumber\\
&+&2zr(\phi^t-\phi^s)+4r\phi^s+(\eta-1)(\phi^t+\phi^s)r_c]+\phi^0[(r^2-1)((r^2-1)z-rr_c+1)
\nonumber\\
&-&\eta((r^2-1)^2z+2r^2-2rr_c-1)]E_{1ef}(t_{1e})h_{1e}(x_3,z,b_3,b)S_t( z)+\phi^0[\eta^2(r^2-1)
\nonumber\\
&+&\eta(1-r^4)+x_3((\eta-1)^2(1-r^2))+\eta (1-r^4)]
\nonumber\\
&+&+\phi^s[2r(r^2-1-\eta-x_3(\eta-1)) ]
\sqrt{\eta(1-r^2)}\big] E_{1ef}(t_{1f})h_{1f}(x_3,z,b_3,b)S_t(x_3)\big\},
\\
M_{AK}&=&32/\sqrt{6}\pi C_F m_B^4 \int dx_B dz dx_3 \int b_B db_B b_3 db_3 \phi_B(x_B,b_B) \phi_D \big\{\big[r^4\phi^0(x_3+x_B-1
\nonumber\\
&-& \eta(x_3+z-2))+r^2\phi^0(\eta^2(x_3+z-2)-\eta(x_3+x_B)+1)+r\sqrt{\eta(1-r^2)}
\nonumber\\
&\times&[\phi^s(\eta-\eta x_3+x_3+x_B-z+3)-\phi^t(\eta-\eta x_3+x_3+xB+z)]+r^3\sqrt{\eta(1-r^2)}
\nonumber\\
&\times&(\phi^s+\phi^t)-(\eta-1)\phi^0(\eta(x_3+z-1)-x_3-x_B) \big]E_{1gh}(t_{1g})h_{1g}(x_B,z,x_3,b_B,b_3)
\nonumber\\
&+&\big[\phi^0(r^2-\eta-1)\big(\eta+r^2(\eta(2x_3+z-2)-2x_3+x_B-z+1)-\eta z+z-1 \big)
\nonumber\\
&+&r\sqrt{\eta(1-r^2)}\big(\phi^s(r^2(1-z)+\eta(x_3-1)-x_3+x_B+z-1)
\nonumber\\
&+&\phi^t(r^2(z-1)+\eta(x_3-1)-x_3+x_B-z+1)\big)\big]
E_{1gh}(t_{1h})h_{1h}(x_B,z,x_3,b_B,b_3)\big\},
\\
F_{TD}&=&8\pi C_F m_B^4F_{K\pi}(s)/\mu_s \int dx_B dx_3 \int b_B db_B b_3 db_3\phi_B(x_B,b_B) \phi_D \big\{(1+r)\big[\eta^2(r-1)x_3
\nonumber\\
&+&\eta(2(r-1)^2x_3-2r+1)+r(x_3(3-2r)+r)-x3-1 \big]E_{2ab}(t_{2a})h_{2a}(x_B,x_3,b_B,b_3)S_t(x_3)
\nonumber\\
&+&\big[(\eta-1)r^4+2r^3(1-2\eta+r_c)-r^2(\eta^2-2\eta r_c+r_c-1)+(\eta-1)(\eta x_3-r_c)
\nonumber\\
&-&2r(\eta(r_c-x_B-1)+r_c+1)\big] E_{2ab}(t_{2b})h_{2b}(x_B,x_3,b_B,b_3)S_t(x_B)\big\}, \label{FTD}
\\
M_{TD}&=&32/\sqrt{6}\pi C_F m_B^4 \int dx_B dz dx_3 \int b_B db_B b db \phi_B(x_B,b_B) \phi_D \phi^0\big\{\big[\eta^2(r^2(2-z-x_3)
\nonumber\\
&+&x_B+z-1)(r^2-1)(r^2(x_3+z-1)-rx_3-x_B-z+1)
\nonumber\\
&+&\eta r(r(r(\eta-1)(x_3+z-2)-x_B-z+2)+x_3+x_B+z-2)\big]
\nonumber\\
&\times& E_{2cd}(t_{2c})h_{2c}(x_B,z,x_3,b_B,b) + \big[(r-1)(\eta+(2\eta-1)r-1)(z(r^2-1)+x_B)
\nonumber\\
&-&x_3(1-\eta)(1-\eta+r(r(2\eta+r-1)-1))\big] E_{2cd}(t_{2d})h_{2d}(x_B,z,x_3,b_B,b)\big\},
\\
F_{AD}&=&8\pi C_F m_B^4 f_B \int dz dx_3\int b db b_3 db_3 \phi_D \big\{\big[2r\phi^s\sqrt{\eta(1-r^2)}(x_3(\eta-1)-2)
\nonumber\\
&+&\phi^0(\eta+-r^2(2\eta+(\eta-1)^2x_3-1)+(\eta-2)\eta x_3+x_3-1)\big]
E_{2ef}(t_{2e})h_{2e}(z,x_3,b_3,b)S_t(x_3)
\nonumber\\
&+& \big[r^4\phi^0(\eta-\eta z+z-1)+
r^2\phi^0(\eta-1)(2z-1-\eta)+2r\sqrt{\eta(1-r^2)}(\phi^s(1+z-\eta)
\nonumber\\
&+&\phi^t(\eta+z-1))-2r^3(z-1)\sqrt{\eta(1-r^2)}(\phi^s+\phi^t)-(\eta-1)z\phi^0-
r^2r_c\sqrt{\eta(1-r^2)}(\phi^s+\phi^t)
\nonumber\\
&-&2r(1+\eta)r_c\phi^0+(1-\eta)r_c\sqrt{\eta(1-r^2)}(\phi^t-\phi^s) \big]E_{2ef}(t_{2f})h_{2f}(z,x_3,b_3,b)
S_t(z)\big\},
\\
M_{AD}&=&32/\sqrt{6}\pi C_F m_B^4 \int dx_B dz dx_3 \int b_B db_B b_3 db_3 \phi_B(x_B,b_B) \phi_D \big\{\big[r^4\phi^0(2(\eta-1)x_3
\nonumber\\
&+&\eta(z-2)-z+1)-r^2\phi^0(\eta^2(x_3+z-2)-x_3+\eta(x_B+z)-x_B-2z+1)
\nonumber\\
&-&r\sqrt{\eta(1-r^2)}(\phi^s(\eta(x_3-1)-x_3+x_B+z+3)+\phi^t(\eta(1-x_3)+x_3+x_B+z-1))
\nonumber\\
&+&r^3 \sqrt{\eta(1-r^2)}(z-1)(\phi^s+\phi^t)+(\eta-1)\phi^0(\eta(x_B+z-1)+x_B+z)\big]
\nonumber\\
&\times&E_{2gh}(t_{2g})h_{2g}(x_B,z,x_3,b_B,b)+\big[\phi^0(\eta-r^2-1)(\eta+x_3-1-\eta(x_3-x_B+z)
\nonumber\\
&+&r^2(\eta(2x_3+z-2)-x_3+1))+r\sqrt{\eta(1-r^2)}\big(\phi^t(r^2(z-1)+\eta(x_3-1)-x_3
\nonumber\\
&+&x_B-z+1)-\phi^s(\eta+r^2(z-1)-\eta x_3+x_3+x_B-z-1)\big)\big]
\nonumber\\
&\times& E_{2gh}(t_{2h})h_{2h}(x_B,z,x_3,b_B,b)\big\},
\end{eqnarray}
where the hard functions are written as
\begin{eqnarray}
h_i(x_1,x_2,(x_3,)b_1,b_2)&=&h_1(\beta,b_2)\times h_2(\alpha,b_1,b_2)  \nonumber\\
h_1(\beta,b_2)&=&\left\{ \begin{array}{ll} K_0(\sqrt{\beta} b_2), \quad &\beta \geq 0, \\
              \frac{i\pi}{2}H^{(1)}_0(\sqrt{-\beta} b_2), \quad &\beta < 0,\end{array} \right.  \nonumber\\
h_2(\alpha,b1,b_2)&=& \left\{ \begin{array}{ll} \theta(b_2-b_1) K_0(\sqrt{\alpha}b_2)I_0(\sqrt{\alpha}b_1), \quad &\alpha \geq 0, \\
              \theta(b_2-b_1) \frac{i\pi}{2}H^{(1)}_0(\sqrt{-\alpha}b_2)J_0(\sqrt{-\alpha}b_1), \quad &\alpha < 0,\end{array} \right.
\label{3-bodyhardfunction}
\end{eqnarray}
where $E_{1mn},E_{2mn}$($m=a,c,e,g$ and $n=b,d,f,h$) are the evolution factors, which given by
\begin{eqnarray}
E_{1ab}(t)&=&\alpha(t)exp[-S_B(t)-S_K(t)],   \nonumber\\
E_{1cd}(t)&=&\alpha(t)exp[-S_B(t)-S_K(t)-S_D(t)]_{b=b_B},  \nonumber\\
E_{1ef}(t)&=&\alpha(t)exp[-S_D(t)-S_K(t)],   \nonumber\\
E_{1gh}(t)&=&\alpha(t)exp[-S_B(t)-S_K(t)-S_D(t)]_{b=b_3}, \nonumber\\
E_{2ab}(t)&=&\alpha(t)exp[-S_B(t)-S_D(t)], \nonumber\\
E_{2cd}(t)&=&\alpha(t)exp[-S_B(t)-S_K(t)-S_D(t)]_{b_3=b_B},  \nonumber\\
E_{2ef}(t)&=&E_{1ef}(t),  \nonumber\\
E_{2gh}(t)&=&E_{1gh}(t),
\end{eqnarray}
in which the Sudakov exponents $S_{(B,K,D)}(t)$ are defined as
\begin{eqnarray}
S_B(t)&=&s\big(\frac{x_B m_B}{\sqrt2},b_B\big)+\frac{5}{3}\int^{t}_{1/b_B}\frac{d\bar{\mu}}{\bar{\mu}}\gamma_q(\alpha_s(\bar{\mu})), \nonumber\\
S_K(t)&=&s\big(\frac{z(1-r^2) m_B}{\sqrt2},b\big)+s\big(\frac{(1-z)(1-r^2) m_B}{\sqrt2},b\big)+
       2\int^{t}_{1/b}\frac{d\bar{\mu}}{\bar{\mu}}\gamma_q(\alpha_s(\bar{\mu})),\nonumber\\
S_D(t)&=&s\big(\frac{x_3 m_B}{\sqrt2},b_3\big)+
       2\int^{t}_{1/b_3}\frac{d\bar{\mu}}{\bar{\mu}}\gamma_q(\alpha_s(\bar{\mu})),
\end{eqnarray}
where the quark anomalous dimension $\gamma_q =-\alpha_s/\pi$. The explicit form for $s(Q,b)$ at one loop can be found in~\cite{Ali:2007ff}. $t_{1x}$ and $t_{2x}$($x=a,b\cdots h$) are hard scales which are chosen to be the maximum of the virtuality of the internal momentum transition in the hard amplitudes as
\begin{eqnarray}
t_{1a}&=& Max \big\{\sqrt{\vert \alpha_{1a} \vert},\sqrt{\vert\beta_{1a}\vert},1/b_B,1/b\big\},\nonumber\\
t_{1b}&=& Max \big\{\sqrt{\vert\alpha_{1b}\vert},\sqrt{\vert \beta_{1b} \vert},1/b_B,1/b\big\},\nonumber\\
t_{1c}&=& Max \big\{\sqrt{\vert\alpha_{1c}\vert},\sqrt{\vert \beta_{1c} \vert},1/b_B,1/b_3\big\},\nonumber\\
t_{1c}^{\prime}&=& Max \big\{\sqrt{\vert\alpha_{1c}\vert},\sqrt{\vert \beta^{\prime}_{1c} \vert},1/b_B,1/b_3\big\},\nonumber\\
t_{1d}&=& Max \big\{\sqrt{\vert\alpha_{1d}\vert},\sqrt{\vert \beta_{1d} \vert},1/b_B,1/b_3\big\},\nonumber\\
t_{1d}^{\prime}&=& Max \big\{\sqrt{\vert\alpha_{1d}\vert},\sqrt{\vert \beta^{\prime}_{1d} \vert},1/b_B,1/b_3\big\},\nonumber\\
t_{2a}&=& Max \big\{\sqrt{\vert\alpha_{2a}\vert},\sqrt{\vert \beta_{2a} \vert},1/b_B,1/b_3\big\},\nonumber\\
t_{2b}&=& Max \big\{\sqrt{\vert\alpha_{2b}\vert},\sqrt{\vert \beta_{2b} \vert},1/b_B,1/b_3\big\},\nonumber\\
t_{2c}&=& Max \big\{\sqrt{\vert\alpha_{2c}\vert},\sqrt{\vert \beta_{2c} \vert},1/b_B,1/b\big\},\nonumber\\
t_{2d}&=& Max \big\{\sqrt{\vert\alpha_{2d}\vert},\sqrt{\vert \beta_{2d} \vert},1/b_B,1/b\big\},\nonumber\\
t_{2e}&=& Max \big\{\sqrt{\vert\alpha_{2e}\vert},\sqrt{\vert \beta_{2e} \vert},1/b_3,1/b\big\},\nonumber\\
t_{2f}&=& Max \big\{\sqrt{\vert\alpha_{2f}\vert},\sqrt{\vert \beta_{2f} \vert},1/b_3,1/b\big\},\nonumber\\
t_{2g}&=& Max \big\{\sqrt{\vert\alpha_{2g}\vert},\sqrt{\vert \beta_{2g} \vert},1/b_B,1/b_3\big\},\nonumber\\
t_{2h}&=& Max \big\{\sqrt{\vert\alpha_{2h}\vert},\sqrt{\vert \beta_{2h} \vert},1/b_B,1/b_3\big\},
\end{eqnarray}
where we have
\begin{eqnarray}
\alpha_{1a}&=&z(1-r^2)m_B^2, \nonumber\\
\beta_{1a}&=&x_B z (1-r^2)m_B^2=\beta_{1b}=\alpha_{1c}=\alpha_{1d}=\alpha^{\prime}_{1c}=\alpha^{\prime}_{1d}, \nonumber\\
\alpha_{1b}&=&(1-r^2)(x_B-\eta)m_B^2,      \nonumber\\
\beta_{1c}&=&-[z(1-r^2)+r^2][(1-\eta)(1-x_3)-x_B]m_B^2, \nonumber\\
\beta_{1d}&=&\{r_c^2-[z(1-r^2)][(1-\eta)x_3-x_B]\}m_B^2, \nonumber\\
\beta^{\prime}_{1c}&=&\{r_c^2-[z(1-r^2)+r^2][(1-\eta)(1-x_3)-x_B]\}m_B^2, \nonumber\\
\beta^{\prime}_{1d}&=&-[z(1-r^2)][(1-\eta)x_3-x_B]m_B^2, \nonumber\\
\alpha_{1e}&=&-[1-z(1-r^2)-r_c^2]m_B^2,                   \nonumber\\
\beta_{1e}&=&-[(1-r^2)(1-z)][\eta+(1-\eta)x_3]m_B^2=\beta_{1f}=\alpha_{1g}=\alpha_{1h} , \nonumber\\
\alpha_{1f}&=&-(1-r^2)(\eta+(1-\eta)x_3)m_B^2  ,               \nonumber\\
\beta_{1g}&=&\{1-[z(1-r^2)+r^2][(1-\eta)(1-x_3)-x_B]\}m_B^2, \nonumber\\
\beta_{1h}&=&-[(1-z)(1-r^2)][(1-\eta)x_3+\eta-x_B]m_B^2, \nonumber\\
\alpha_{2a}&=&x_3(1-\eta)m_B^2 ,\nonumber\\
\beta_{2a}&=&x_3x_B(1-\eta)m_B^2=\beta_{2b}=\alpha_{2c}=\alpha_{2d}, \nonumber\\
\alpha_{2b}&=&x_B(1-\eta)m_B^2 ,     \nonumber\\
\beta_{2c}&=&-[(1-r^2)(1-z)-x_B][\eta+(1-\eta)x_3]m_B^2 ,\nonumber\\
\beta_{2d}&=&-(1-\eta)x_3[(1-r^2)z-x_B]m_B^2, \nonumber\\
\alpha_{2e}&=&-[1-x_3(1-\eta)]m_B^2  ,               \nonumber\\
\beta_{2e}&=&-[r^2+z(1-r^2)](1-\eta)(1-x_3)m_B^2=\beta_{2f}=\alpha_{2g}=\alpha_{2h} , \nonumber\\
\alpha_{2f}&=&\{r_c^2-[r^2+z(1-r^2)](1-\eta)\}m_B^2 ,                \nonumber\\
\beta_{2g}&=&\{1-[(1-r^2)(1-z)-x_B][\eta+(1-\eta)x_3]\}m_B^2, \nonumber\\
\beta_{2h}&=&-[r^2+z(1-r^2)-x_B](1-\eta)(1-x_3)m_B^2.
\end{eqnarray}

\end{document}